\documentstyle[fleqn]{article}
\newcommand{\bc}{\begin{center}}
\newcommand{\ec}{\end{center}}
\newcommand{\bq}{\begin{quotation}}
\newcommand{\eq}{\end{quotation}}
\newcommand{\beq}{\begin{equation}}
\newcommand{\eeq}{\end{equation}}
\newcommand{\la}{\label}
\newcommand{\un}{\underline}
\newcommand{\ov}{\overline}
\newcommand{\dd}{\mbox{d}}

\title{Minimal Information in Velocity Space}
\author{G. Evrard\thanks{Astronomy Unit,
 School of Mathematical Sciences,
 Queen Mary and Westfield College,
Mile End Road, London E1 4NS, U.K.}\ \thanks{
Groupe de Recherche en Astronomie et Astrophysique du Languedoc,
URA 1368 CNRS/Montpellier II, c.c. 072,
Universit\'e Montpellier II, place Eug\`ene Bataillon,
F-34095 Montpellier Cedex 05, FRANCE
(permanent adress).
Fax: (33) 67 14 45 35.
E-mail: evrard@graal.univ-montp2.fr}}
\begin{document}
\maketitle
\begin{abstract}
Jaynes' transformation group principle can allow the computation of the prior
density functions
describing minimal knowledge for a velocity. The improper prior is uniform in
the unbounded velocity space
of classical mechanics. In the relativistic case, however, it reads $\ov{\mu}
(\beta_x,\beta_y,\beta_z)
=a\times[1-(\beta_x^2+\beta_y^2+\beta_z^2)]^{-2}$, but it can be rewritten as
a uniform volumetric
distribution, when the velocity space is given a non-trivial metric.
\end{abstract}
\section{Introduction}
Physicists seldom escape the need to express a state of incomplete knowledge
of parameters describing a physical system.
When certainty is not at hand, the appropriate
language to describe information is probability. The Bayesian approach to
probability has proved to both be free of the inconsistencies induced
by the classical ``frequentist'' theory, and to provide an elegant
solution of many controversial
problems, particularly in statistical mechanics\cite{garrett}.
However, the necessary
assignment of a prior distribution describing {\em minimal knowledge} can be,
when not neglected, a major difficulty. The search for the prior density
function can lead to appropriate non-trivial results,
even in the most elementary problems, as
pointed out in the following, with the example of the velocity of a particle,
in classical and special relativistic mechanics.
\section{Minimal information and measure}

I hereafter
take the Bayesian viewpoint on probability theory \cite{garrett,loredo},
according to which
the probability is a real-numbered measure of one's belief in the validity
of a logical proposition $A$, given incomplete knowledge $C$. By incomplete,
I mean that $C$ does not allow one to establish the truth or falsehood of $A$
with certainty.
In this scheme, a probability is always conditional, in the
sense that it can be assigned to a proposition $A$, only assuming some previous
knowledge $C$. This is clearly recognized by denoting the probability of
proposition $A$ given information $C$ as $p(A|C)$.

Let $x$ be a parameter taking its value in a given range, and characterizing
the physical system under study ( from a chosen point of view ).
The knowledge $C$ obtained by some observation or experiment is
then usually represented by a probability density that will be noted,
following Tarantola \cite{tarantola}  by an overlined symbol, e.g.~:
\begin{equation}
\la{pdnot}
\ov{f}_{x}:x\rightarrow\ov{f}_x(x)\, .
\eeq
Proposition $A$ can be a statement about the value of $x$ (e.g.:
$A=(x\in]x_1,x_2[)$ ).
The density $\ov{f}_{x}$ allows then the computation of the probability
$p(A|C)$ by the
usual relation~:
\beq
\la{pduse}
p(A|C)=\int_{x_{1}}^{x_{2}}\,\ov{f}_{x}(x) \, \dd x\, .
\eeq
If a different parametrization $y=T(x)$ is used to describe the system,
then the information $C$ induces a probability density $\ov{f}_y$ related to
$\ov{f}_x$ by
\beq
\la{rcp}
\ov{f}_x(x)=\ov{f}_y(y)\,|J_T|\, .
\eeq
where $|J_T|$ stands for the Jacobian of the change of parametrization $T$.
Equation (\ref{rcp}) ensures that the probability of proposition
$A=(x\in]x_1,x_2[)=
\left[y\in T(]x_1,x_2[)\right]\;$ given information $C$ is attributed the same
 value whether the
$x$ or the $y$ parametrization is used. This is hereafter referred to as the
relation of conservation of probabilities.

Now, prior to any observation, complete ignorance itself is also a state of
knowledge, and it should be
described by a density function (d.f.) too. Indeed, the raw definition of
the parameter $x$ and the properties it possesses are inevitably associated
with some
information $I$~: the state of minimal information. The corresponding
d.f. will be noted $\ov{\mu}_x$ and is usually termed
the least informative d.f., or the prior d.f. It must be emphasized
that the form of $\ov{\mu}_x$ is generally not trivial, in particular,
$\ov{\mu}_x$ is not necessarily a constant. Evidence of this is given by
the application of Eq.(\ref{rcp}) to two parametrizations $x$ and $y$,
where minimal knowledge is described by
\beq
\la{tmi}
\ov{\mu}_x(x)=\ov{\mu}_y(y)\,|J_T|\, .
\eeq
Hence the prior $\ov{\mu}$ cannot be constant in, for example,
two parametrizations $x$ and $y$ related by a nonlinear transformation $T$.
As a consequence, a general method of inference of the form of $\ov{\mu}$
must be sought.

Jaynes \cite{jaynes} has proposed a ``transformation group'' method relying on
the basic principle that ``the state of minimal information is described
by the same d.f. in two different parametrizations in which the problem is
equivalently defined'', or, more widely stated by Tarantola and Valette
\cite{tarantolavalette}, ``the least informative d.f. is form-invariant under
the
transformations that leave invariant the equations of physics''. Therefore,
in two parametrizations $x$ and $y$ related by such a transformation
(i.e. $y=T(x)$ ), Jaynes' principle reads~:
\beq
\la{jp}
\ov{\mu}_x\equiv\ov{\mu}_y\, \;\;\;\mbox{(i.e.
$\forall\,y,\;\ov{\mu}_y(y)=\ov{\mu}_x(y)$ )}.
\eeq
Thus, the form of this function is constrained by the relation of
conservation of probabilities (\ref{rcp}), which imposes the constraint
\beq
\la{ip}
\ov{\mu}_{x}(x) \, \dd x=\ov{\mu}_{x}[T(x)]\,\dd [T(x)]\, .
\eeq
This relation must be valid for all $x$ and for every allowed
transformation $T$.

In a later version \cite{tarantola} of his probabilistic approach to inverse
problems,
Tarantola suggests the use of volumetric probabilities $\mu$ instead
of probability densities $\ov{\mu}$. Let $x=(x^1,x^2,\cdots,x^n)$ be
a given parametrization of the physical system under study.
Probability densities $\ov{f}$ have to be
multiplied by the differential element $\dd x=\dd x^1\,\dd x^2\cdots\dd x^n$
 of the coordinates to get
an infinitesimal probability, whereas volumetric probabilities $f$ have to
be multiplied by a volume element $\dd V(x^1,x^2,\cdots,x^n)$
 to get the same equality
\beq
\la{pdvsvp}
\ov{f}_x(x^1,x^2,\cdots,x^n)\,\dd x^1\,\dd x^2\cdots\dd x^n
=f(x^1,x^2,\cdots,x^n)\,\dd V(x^1,x^2,\cdots,x^n) \, .
\eeq
Tarantola chooses then to describe ignorance by a constant volumetric
probability $\mu=\mbox{const.}$ This arbitrary choice corresponds
to the ``natural'' feeling that in the case of minimal knowledge,
the probability is proportional to the volume element\footnote{This is
perhaps
where the automatic use of a constant probability density to describe
ignorance originates, by misusing the differential of the coordinates
as the volume element.}. From this choice, the search for the correct
probability density $\ov{\mu}_x$ turns into the search of the right volume
element $\dd V(x)$. Indeed, Eq. (\ref{pdvsvp}) becomes
\beq
\la{corr1}
\ov{\mu}_x(x^1,x^2,\cdots,x^n)\,\dd x^1\,\dd x^2\cdots\dd x^n=
\dd V(x^1,x^2,\cdots,x^n)\, \times \mbox{const.}
\eeq\
{}From this point of view, Jaynes' principle can be stated as~: ``In two
parametrizations where the laws of physics take the same form, the
volume element has the same form''. In other words, the volume element
is unchanged under a transformation $T$ that ``leave invariant the
laws of physics''~:
\beq
\la{ive}
\dd V(x^1,x^2,\cdots,x^n)=\dd V [T(x^1,x^2,\cdots,x^n)] \, .
\eeq
For a sufficiently well-defined parameter, these conditions should permit
the identification
 of an ``objective\footnote{ in the sense that it takes the same form
whether one uses the $\un{x}$ or the $\un{y}=T(\un{x})$ parametrization.}''
volume element.
 Furthermore, should the parameter
space, where $\un{x}$ is defined, be given a metric, its form can be
constrained.
 Indeed, if the line element
 is
\beq
\la{corb1}
\dd l^2=g_{ij}\,\dd x^i\,\dd x^j \, ,
\eeq
the volume element is given by
the square root of the determinant of the metric $g_{ij}$ multiplied by
the differential of the parameters, that is~:
\beq
\la{corb2}
\dd V(x^1,x^2,\cdots,x^n)=\sqrt{|g_{ij}|}\,\dd x^1\,\dd x^2\cdots \dd x^n\, .
\eeq
We will hereafter use this principle to infer the appropriate least
informative d.f.
for the velocity of a particle with non-zero rest-mass, using both classical
and
relativistic mechanics, and the corresponding norms.\\

\section{Velocity in classical mechanics}

Let $K$ and $K'$ be two inertial reference frames where observers
$O$ and $O'$respectively, are at rest, and such that for each of them, the
 other is moving along the $x$ (or $x'$) axis in his reference frame. Let $P$
be a point-like material particle, moving freely along this common axis. The
velocity of $P$ can be indicated by its value $v_x$ for $O$ and $v_x'$ for
$O'$. Then,
$v_x$ and $v_x'$ are related by the classical law of addition of velocities
\beq
\la{cav}
v_x'=v_x-v_r \, ,
\eeq
where $v_r$ is the relative velocity of $O'$ as seen by $O$.

Let $\ov{\mu}_x$ be the d.f. which represents
minimal knowledge of this velocity for observer $O$, and
$\ov{\mu}_x'$ the corresponding least informative d.f. for the velocity
of $P$ as seen by $O'$. The principle of Galilean relativity states
that neither of the two observers is in a special situation with respect
to the laws of mechanics. Therefore, by Jaynes' principle, both observers $O$
and $O'$ must describe
 minimal knowledge of the particle's velocity by the same d.f.,
that is
\beq
\la{jpvc}
\ov{\mu}_x\equiv\ov{\mu}_x'\, .
\eeq
Now, $v_x$ and $v_x'$ can be seen as two different parametrizations for
the same physical system. The rule of conservation of probabilities (\ref{rcp})
holds, and therefore the d.f. $\ov{\mu}_x$ must obey the equivalent of
Eq.(\ref{ip}), i.e.~:
\beq
\la{cvc}
\ov{\mu}_x(v_x)\,\dd v_x=\ov{\mu}_x(v_x-v_r)\,\dd v_x
\eeq
This relation must be valid for all $v_r$ and all $v_x$. This implies
\beq
\la{dfcv}
\ov{\mu}_x(v_x)=\mbox{const.}
\eeq
Note that, as $-\infty<v_x<+\infty$, this d.f. is an improper
prior\cite{jaynes2} (this is
a common feature of a least informative d.f.).

 In terms of volumetric
 probabilities, a comparison of the result (\ref{dfcv}) and Eq.
(\ref{corr1}) gives immediately the correct volume element~:
\beq
\la{be1d}
\dd V(v_x) \,\propto \, \dd v_x\,.
\eeq
The former case can be extended to 2 and 3 dimensional cases, in which
 the velocities measured by two equivalent observers are related by
\beq
\la{r23d}
\un{v}'=R(\un{v}-\un{v}_r) \, ,
\eeq
where $\un{v}'$, $\un{v}$ and $\un{v}_r$ are 2 or 3-vectors respectively
standing for the velocity of~: $P$ seen by $O'$, $P$ seen by $O$, and
$O'$ seen by $O$. $R$ is, depending on the case, a 2-D or 3-D rotation
matrix.

Since $O$ and $O'$ are equivalent observers, they must describe
minimal knowledge of the velocity by the same d.f. However, as previously,
$\un{v}$ and $\un{v}'$ can be seen as two different parametrizations of the
same system and hence, the rule of conservation of probabilities implies~:
\beq
\la{rcpcv3d}
\ov{\mu}_{\un{v}}(\un{v})\,\dd \un{v}=\ov{\mu}_{\un{v}}\left[
R(\un{v}-\un{v}_r)\right]\, \dd \un{v}'\, .
\eeq
The Jacobian of the transformation (\ref{r23d}) being equal to 1, this
also leads to
\beq
\la{dfcv3d}
\ov{\mu}_{\un{v}}(\un{v})=\mbox{const.}
\eeq
The deduction of the volume element is, as previously, straightforward.
For the 2-D case, we get
\beq
\la{ve2d}
\begin{array}{l}
\dd V(v_x,v_y)=\mbox{const.} \times \dd v_x \, \dd v_y \\
\dd V(v_x,v_z)=\mbox{const.} \times \dd v_x\, \dd v_z\, , \\
\dd V(v_y,v_z)=\mbox{const.} \times \dd v_y\, \dd v_z
\end{array}
\eeq
and for the 3-D case, we get, equivalently,
\beq
\la{ve3d}
\dd V(v_x,v_y,v_z)=\mbox{const.} \times \dd v_x \, \dd v_y \, \dd v_z\, .
\eeq
This allows the computation of an ``objective'' metric in velocity space.
For simplicity's sake, it is convenient to use the polar coordinate
system $(v,\theta,\varphi)$ related to the cartesian coordinates $(v_x,v_y,
v_z)$ by the relations~:
\beq
\la{corc1}
\left\{
\begin{array}{l}
v_x=v\, \cos\theta\\
v_y=v\,\sin\theta\,\sin\varphi\\
v_z=v\,\sin\theta\,\cos\varphi
\end{array}
\right.
\end{equation}
For reasons of isotropy, the length element must take the form~:
\beq
\la{lmel}
\dd l^2=f_1(v)\, \dd v^2+f_2(v)[\dd \theta^2+\sin^2 \theta
\, \dd \varphi^2]\, .
\eeq
Equation (\ref{be1d}) corresponds  to a fixed direction of $\un{v}$, i.e.
$ \theta =0$. This leads to $\dd l^2=
g_{11}^2=$const., say~:
\beq
\la{cl1ve}
g_{11}^2=a^2\, .
\eeq
In polar coordinates, equations (\ref{ve2d}) and (\ref{ve3d}) read
\beq
\la{vpol}
\begin{array}{l}
\dd V(v,\theta)=\mbox{const.}\times v \,\dd v \,\dd \theta \\
\dd V(v, \theta,\varphi)=\mbox{const.} \times v^2 \sin \theta \, \dd v
\, \dd \theta \, \dd \varphi\, .
\end{array}
\eeq
This imposes $f_2(v)=\mbox{const.} \times v^2$, that is
\beq
\la{cl2ve}
\begin{array}{l}
g_{22}=b^2 \, v^2\\
g_{33}=b^2\,v^2 \,\sin^2\theta
\end{array}
\eeq
 and again, symmetry considerations impose $a=b$. Thus the invariant metric
of the velocity space in classical mechanics takes the Euclidean form
\beq
\la{cmm}
\begin{array}{ll}
\dd l^2 & = a^2\,[\dd v^2+v^2\,(\dd \theta^2+ \sin ^2 \theta\,
\dd \varphi ^2)]
\\
         & = a^2\,(\dd v_x^2+\dd v_y^2 + \dd v_z^2)\, .
\end{array}
\eeq
This length element, together with the choice of a constant volumetric
probability to describe
minimal information,
ensures that the prior for the velocity expressed in cartesian coordinates
is constant,
in one, two
and three dimensional cases, as required by Eqs. (\ref{dfcv}) and
(\ref{dfcv3d}).

\section{Velocity in special relativity}
We now express the velocities in terms of their ratio to the
speed of light (i.e. : $\beta=v/c$). The difference from the former section is
that, instead of the Galilean laws for the addition of velocities, the
relativistic laws have now to be applied.

In the one-dimensional case,
the relativistic equivalent of Eq. (\ref{cav}) reads
\beq
\la{rav}
\beta'=\frac{\beta-\alpha}{1-\alpha \,\beta}\, ,
\eeq
where $\beta'=v'/c$ and $\alpha=v_r/c$. It is convenient to use the
parametrization defined by $b= \mbox{arctanh}\,\beta$, $a= \mbox{arctanh}
\,\alpha$ and
$b'=\mbox{arctanh}\,\beta'$, in which Eq.(\ref{rav}) takes exactly
the same form as Eq.(\ref{cav}), that is~:
\beq
\la{rav2}
b'=b-a\, .
\eeq
Consequently, Jaynes' principle leads to~:
\beq
\la{rdf1}
\ov{\mu}_{b}(b)=\mbox{const.}
\eeq
Back in the $\beta$ parametrization (using Eq. \ref{rcp} ), this becomes
\beq
\la{rdf2}
\ov{\mu}_{\beta}(\beta)=\frac{\mbox{const.}}{1-\beta^2}\, .
\eeq
Note that as $-1<\beta<1$, this d.f. is not normalizable, cf Eq. (\ref{dfcv}).

For the two and three dimensional cases, the relativistic law of
addition of velocities reads~:
\beq
\la{rav3d}
\left\{
\begin{array}{l}
\beta_x'=(\beta_x-\alpha)/(1-\alpha\,\beta_x)\\
\beta_y'=\beta_y\, (1-\alpha^2)^{1/2}/(1-\alpha\,\beta_x)\\
\beta_z'=\beta_z\, (1-\alpha^2)^{1/2}/(1-\alpha\,\beta_x)
\end{array}
\right.
\eeq
It is more convenient to use the parametrization $(\beta,\cos \theta,\varphi)$
 related to $(\beta_x,\beta_y,\beta_z)$ by
\beq
\la{cpb}
\left\{
\begin{array}{l}
\beta_x=\beta\,\cos \theta \\
\beta_y=\beta \, \sin \theta \, \sin \varphi \\
\beta_z= \beta \, \sin \theta \, \cos \varphi
\end{array}
\right.
\eeq
and similarly for the primed reference frame. These equations lead, after
some calculation, to the following relations between
 $(\beta,\cos\theta,\varphi)$ and $(\beta',\cos\theta',\varphi')$
\beq
\la{cpb2}
T_{\alpha}:\;
\left(
\begin{array}{c}
\beta\\
\cos \theta\\
\varphi
\end{array}
\right)
\rightarrow
\left(
\begin{array}{ll}
\beta' & = [(1-\alpha \,\beta\,\cos \theta)^2-(1-\beta^2)(1-\alpha^2)]^{1/2}
(1-\alpha \,\beta\, \cos\theta)^{-1}\\
\cos \theta' & = (\beta\,\cos\theta-\alpha)
[(1-\alpha \,\beta\,\cos \theta)^2-(1-\beta^2)(1-\alpha^2)]^{-1/2}\\
\varphi' & = \varphi
\end{array}
\right)
\eeq
One can check that the Jacobian of such a transformation allows
the writing of a  simple equality~:
\beq
\la{jtr}
\frac{\beta'^2}{(1-\beta'^2)^2}\,\dd \beta'\,\dd (\cos\theta')\,\dd\varphi'=
\frac{\beta^2}{(1-\beta^2)^2}\,\dd\beta\,\dd (\cos\theta )\,\dd\varphi \, .
\eeq
The equivalent of Eq.(\ref{ip}), expressing both Jaynes' principle and the
principle of special relativity reads
\beq
\la{mrp}
\ov{\mu}_{\beta\cos\theta\varphi}(\beta',\cos \theta',\varphi')\,\dd \beta'\,
\dd(\cos\theta')\,
\dd \varphi'=\ov{\mu}_{\beta\cos\theta\varphi}(\beta,\cos \theta,\varphi)\,
\dd\beta\,\dd(\cos\theta)\,
\dd\varphi\, ,
\eeq
which means that two observers $O$ and $O'$ at rest in two inertial frames
$K$ and $K'$ moving with a relative velocity $\alpha$, describe the
minimal information about a particle's velocity $(\beta,\cos\theta,\varphi)$
by the same d.f. $\ov{\mu}_{\beta\cos\theta\varphi}$. Comparison of Eqs.
(\ref{jtr}) and (\ref{mrp})
yields an obvious solution for $\ov{\mu}_{\beta\cos\theta\varphi}
(\beta,\cos\theta,\varphi)$, which reads~:
\beq
\la{dfsol}
\ov{\mu}_{\beta\cos\theta\varphi}(\beta,\cos\theta,\varphi)=\frac{\beta^2}
{(1-\beta^2)^2}\times a
\eeq
where $a$ is a constant.

We can use isotropy arguments to assert that the corresponding metric of
the parameter space must take the form
\beq
\la{bmet}
\dd l^2=g_1(\beta)\,\dd\beta^2+g_2(\beta)\left[ \frac{\dd (\cos\theta)^2}{
\sin^2 \theta}+\sin^2\theta\,\dd\varphi^2\right]\, .
\eeq
So that we have from (\ref{bmet}) and (\ref{rdf2}), for the fixed, known
direction of the velocity $\theta=0$,
$\;\dd\theta=0$
\beq
\la{mft}
\left[g_1(\beta)\right]^{1/2}=\frac{a}{1-\beta^2} \, .
\eeq
The more general three-dimensional case similarly leads from (\ref{bmet})
and (\ref{mft}) to
\beq
\la{mst}
\left[g_1(\beta)\,g_2(\beta)^2\right]^{1/2}=
b\times\frac{\beta^2}{(1-\beta^2)^2
} \, .
\eeq
We can thus infer the form of the two functions $g_1$ and $g_2$. Imposing that
when $\beta$ tends towards 0 (i.e. the velocity of the particle is low
compared to $c$), the metric takes the classical form (\ref{cmm}),
it is straightforward to obtain
\beq
\la{rmetric}
\dd l^2 =a^2\,\left[
\frac{1}{(1-\beta^2 )^2}\,\dd \beta^2 +\frac{\beta^2}{(1-\beta^2)}(
\dd\theta^2 +\sin^2 \theta\,\dd\varphi^2 )
\right] \, .
\eeq
It is then purely a technical matter to get the corresponding form of
the metric tensor
for the Cartesian $(\beta_x,\beta_y,\beta_z)$ coordinates. The result is
\beq
\la{ccrmet}
\left[ g_{ij}\right] =\frac{a^2}
{[1-(\beta_x^2+\beta_y^2+\beta_z^2)]^2}\,
\left[
\begin{array}{ccc}
1-\beta_y^2-\beta_z^2 & \beta_x\,\beta_y & \beta_x\, \beta_z\\
 \beta_x\,\beta_y & 1-\beta_x^2-\beta_z^2 & \beta_y\,\beta_z\\
 \beta_x\, \beta_z &\beta_y\,\beta_z & 1-\beta_x^2-\beta_y^2
\end{array}
\right]\, .
\eeq
This metric allows the direct computation of the prior for a velocity
in one, two, and three dimensional cases. In particular, both Eqs.
(\ref{dfsol})
and (\ref{ccrmet}) yield~:
\beq
\la{modif}
\ov{\mu}_{\beta_x \beta_y \beta_z}(\beta_x,\beta_y,\beta_z)=\frac{a}{
[1-(\beta_x^2+\beta_y^2+\beta_z^2)]^2}
\eeq

\section{Comments}
The volume element induced by Eq. (\ref{rmetric}) leads to an infinite volume
for the whole velocity space $-1<\beta<1$, $0\leq\theta\leq\pi$, $0\leq
\varphi < 2\,\pi$. This is also related to the non-normalizability of the d.f.
$\ov{\mu}_{\beta\cos\theta\varphi}$ inferred in Eq. (\ref{dfsol}). As $\beta$
tends towards 1, the ``distance'' between two ``close'' points $l([\beta,
\beta+\dd\beta])$ tends to infinity. This should have been expected, since the
equivalent classical case is $v\rightarrow +\infty$. The length element
(\ref{rmetric})
also implies that in velocity space, the invariant ``distance'' of a point
 to the sphere
$\beta=1$ is infinite, in any reference frame,therefore, the statement that
a velocity $v$
is ``close'' to $c$ is meaningless. An other consequence
is, that choosing $0.95 \,c$ as an ``average'' value for $v$,
when one has the information $0.9\,c\leq v \leq c$, appears logically nonsense.
In any case, in the energy parametrization $E=E_0\,(1-\beta^2)^{-1/2}$, the
corresponding knowledge would only give a lower value for the energy of the
particle, and therefore, it would not come to one's mind to try to summarize
this information in terms of one ``average'' value.

The metric (\ref{rmetric}) and (\ref{ccrmet}) is ``objective'' in the sense
that it defines a measure element in the velocity space which is invariant
under any Lorentz velocity-transformation, accordingly to the special principle
of relativity. Both the metric (\ref{ccrmet}) and the densities (\ref{rdf2})
or (\ref{modif})
deduced from the special principle of relativity yield to the metric
(\ref{cmm})
and the densities (\ref{dfcv}) and (\ref{dfcv3d}) respectively, when the
classical approximation
$c\rightarrow+\infty$ is made.

It might appear strange that $\ov{\mu}_{xyz}\neq\ov{\mu}_x\,\ov{\mu}_y\,
\ov{\mu}_z$. Indeed, it is a counter-example of the intuitive general equality
postulated by Tarantola \cite{albert}. Actually, in the Galilean case, the
equality
holds, from (\ref{dfcv}) and (\ref{dfcv3d}). Yet, for the relativistic case,
it is clear that it is not valid anymore. Indeed, the appearance of transverse
terms in the metric (\ref{ccrmet}) makes minimal information on the velocity
in one fixed and known direction depend on the orthogonal velocity. This is a
direct consequence of the relativity of time, and should not therefore be
surprising. It illustrates however, how important it is to clearly define the
problem
 before any prior is to be inferred by Jaynes' method. In other
words, the parameters have to be precisely characterized, and the invariance
 transformation groups and subgroups found, previously to the application
of the transformation-group method.\\

To summarize, the prior density functions for a velocity
have been found in 1, 2 and 3 dimensional cases, in Galilean and special
relativity, using Jaynes' transformation group method. The inference
of the corresponding invariant ``objective'' measure in the velocity space
illustrates the relations between measure, volume element and
least-informative density function.
Even in this very simple case, the result (i.e. Eqs. (\ref{ccrmet}) and
(\ref{modif}) )
is non-trivial. As the assignment of
prior probabilities is a necessary first step before any consistent (i.e.
Bayesian) probability-based inference method is put into practice, this shows
that the precise definition of the parameters must be given one's full
attention.
In less obvious cases, this task is more dificult, but it can also lead to
important conclusions. An application of this technique to derive the
minimal information description of {\em cosmological} parameters is underway,
and appears very promising.\\

\noindent {\bf Acknowledgements}\\
I wish to thank Albert Tarantola for friendly discussions and encouragements.
I am also grateful to Peter Coles for precious help and comments.\\
This work was partly supported by the European Community HCM Network,
contract number ERB CHRX-CT93-0129.


\begin{thebibliography}{99}
\bibitem{garrett}A.J.M. Garrett in: Maximum entropy in Action, eds.
B. Buck and V.A. Macaulay (Oxford University Press, 1991), pp.139-170.
\bibitem{loredo}T.J. Loredo, in:  Maximum entropy and
 Bayesian Methods, ed. P.F. Foug\`ere
 (Kluwer Academic Publishers, Dordrecht, The Netherlands
1990), pp. 81-142.
\bibitem{tarantola}A. Tarantola, in:
 Les Houches Summer School Lectures in Theoretical Physics,
Session L (1988): Oceanographic and Geophysical Tomography,
eds. Y. Desaubies, A. Tarantola  and J. Zinn-Justin
(North-Holland Publishing Co., Amsterdam, 1989), pp. 3-27
\bibitem{jaynes}E.T. Jaynes, IEEE Transactions on Systems Science and
Cybernetics,
Vol SSC-4, No 3 (1968) 227.\\
Reprinted in:  E.T. Jaynes: Papers on Probability, Statistics and Statistical
Physics,
ed. R.D. Rosenkrantz
(Kluwer Academic Publishers, Dordrecht, The Netherlands, 1983, reprinted 1989)
\bibitem{jaynes2} E.T. Jaynes in: Bayesian Analysis in Econometrics and
Statistics, ed. A Zellner
(North Holland Publishing Co., Amsterdam 1980).
Reprinted as the above reference.
\bibitem{tarantolavalette}A. Tarantola and B. Valette, J. Geophys. 50 (1982)
159.
\bibitem{albert} A. Tarantola, Inverse Problem Theory: Methods for Data Fitting
and Model Parameter Estimation (Elsevier, Amsterdam, 1987)
\end{thebibliography}
\end{document}